\documentclass[11pt,a4paper]{article}
\usepackage{fullpage}
\usepackage{amssymb,amsmath}
\usepackage[round]{natbib}
\usepackage{lineno}
\usepackage{amsfonts,amsbsy}
\usepackage{graphicx}
\usepackage{color}

\def\e{\mathrm{e}}
\def\i{\mathrm{i}}
\def\d{\mathrm{d}}

\def\beq{\begin{equation}}
\def\eeq{\end{equation}}

\def\bx{\boldsymbol{x}}

\def\bxi{\boldsymbol{\xi}}

\def\cc{\mathrm{c.c.}}

\newcommand{\av}[1]{\langle #1 \rangle}
\newcommand{\dpar}[2]{\frac{\partial #1}{\partial #2}}

\newcommand{\dt}[2]{\frac{\mathrm{d} #1}{\mathrm{d} #2}}

\newcommand{\eqn}[1]{(\ref{eqn:#1})}
\newcommand{\lab}[1]{\label{eqn:#1}}
\newcommand{\inter}[1]{\quad \textrm{#1} \quad}

\newcommand{\su}[1]{^{\left(#1\right)}}
\newcommand{\bs}[1]{\boldsymbol{#1}}

\def\XXint#1#2#3{{\setbox0=\hbox{$#1{#2#3}{\int}$}
\vcenter{\hbox{$#2#3$}}\kern-.5\wd0}}

\title{Deriving the Young--Ben Jelloul model of near-inertial waves by Whitham averaging}
\author{J. Vanneste \smallskip \\
\normalsize{School of Mathematics and Maxwell Institute for Mathematical Sciences}, \\ \normalsize{University of Edinburgh, Edinburgh EH9 3FD, UK}}
\date{\today}

\begin{document}
\maketitle

\begin{center}
\begin{minipage}{0.8\textwidth}
Oceanic near-inertial waves -- internal waves with frequencies close to the local Coriolis frequency $f_0$ -- are strongly influenced by the presence of mean currents. To study this influence, \citet{youn-benj} derived an asymptotic model that describes the slow modulation of the amplitude of these waves about their rapid oscillation at frequency $f_0$. Here we show that this model can be obtained within a variational framework, by (Whitham) averaging the Lagrangian of the hydrostatic--Boussinesq equations over the wave period $2\pi/f_0$. The derivation leads to a variational formulation of the Young--Ben Jelloul model from which its conservation laws can be obtained systematically.

\bigskip
\hrule
\end{minipage} 
\end{center}
\bigskip

\noindent
In this note, we derive the \citeauthor{youn-benj} model (\citeyear{youn-benj}, hereafter YBJ) by applying \citet{whit74} averaging to a variational form of the hydrostatic--Boussinesq equations on the $\beta$-plane. These equations are obtained from the variational principle
\[
\delta \int L \, \d t = 0
\]
for the Lagrangian 
\[
L = \int \left(\frac{1}{2} \left({\dot x}^2 + {\dot y}^2 \right) - \left(f_0 y + \frac{\beta y^2}{2}\right) \dot x + \theta z + p \left( \left| \dpar{\bx}{\bs{a}} \right| - 1 \right) \right) \, \d \bs{a}
\]
and independent variations of the positions $\bx=(x,y,z)$ (regarded as functions labels $\bs{a}$ and time) and Lagrange multiplier (pressure) $p$ \citep[see][]{salm88b,salm98}. In the above, the buoyancy $\theta$ is a fixed function of the labels which is therefore materially conserved: $\dot \theta=0$. 

Consider a background flow, with flow map $\bx = \bs{X}(\bs{a},t)$ satisfying $|\partial\bs{X}/\partial\bs{a}|=1$, pressure $p(\bs{a},t)$ and buoyancy satisfying $\partial_z p = \theta$. When this flow is perturbed, the particle positions can be written as
\beq \lab{glm}
\bs{x}(\bs{a},t)=\bs{X}(\bs{a},t)+\bs{\xi}(\bs{X}(\bs{a},t),t),
\eeq
using a generalised-Lagrangian-mean formulation  \citep[e.g.][]{buhl09}. Denoting the background pressure in Eulerian form by $P(\bs{X}(\bs{a},t),t)$, we write the total pressure as
\begin{eqnarray}
p(\bs{a},t)&=& P(\bs{X}(\bs{a},t) +\bs{\xi}(\bs{a},t),t) + q(\bs{X}(\bs{a},t),t) \nonumber \\
&=&  P(\bs{X}(\bs{a},t),t) +\bs{\xi}(\bs{a},t) \cdot \nabla P(\bs{X}(\bs{a},t),t) + q(\bs{X}(\bs{a},t),t)+\cdots, \lab{pressure}
\end{eqnarray}
where the displacement $\bxi$ is introduced inside $P$ for convenience, as part of the definition of the perturbation pressure  $q$ \citep[cf.][]{beck-grim}. The second line omits quadratic and higher-order terms in the perturbation which can be neglected. 

We now introduce \eqn{glm}--\eqn{pressure} into the Lagrangian and expand in powers of the perturbation. Assuming that the background flow is a solution of the hydrostatic--Boussinesq equations, the first-order term vanishes (since the action is stationary for the background flow). To evaluate the next-order, quadratic term, we note that
\begin{eqnarray*}
\left| \dpar{\bs{x}}{\bs{a}} \right| &=& 1 + \nabla \cdot \bs{\xi} + \dpar{(\xi,\eta)}{(x,y)} + \dpar{(\eta,\zeta)}{(y,z)} + \dpar{(\zeta,\xi)}{(z,x)} + \cdots \\
&=&  1 + \nabla \cdot \bs{\xi} + \tfrac{1}{2} \nabla \cdot \left( (\nabla \cdot \bxi) \bxi - (\bxi \cdot \nabla) \bxi \right) + \cdots,
\end{eqnarray*}
and that the two contributions of the background pressure simplify according to
\begin{eqnarray*}
&\tfrac{1}{2} P \nabla \cdot \left( (\nabla \cdot \bxi) \bxi - (\bxi \cdot \nabla) \bxi \right) + \bxi \cdot \nabla P \, \nabla \cdot \bxi = & \\
& \tfrac{1}{2} \nabla P \cdot \left( (\nabla \cdot \bxi) \bxi + (\bxi \cdot \nabla) \bxi \right) + \nabla \cdot ( \cdots) = - \tfrac{1}{2} \bxi \cdot \nabla \nabla P \cdot \bxi + \nabla \cdot ( \cdots), &
\end{eqnarray*}
where $\nabla \nabla P$ denotes the Hessian of $P$ and $\nabla \cdot ( \cdots)$ a divergence term does not contribute to the Lagrangian. We then obtain the quadratic Lagrangian
\beq \lab{L2}
L\su{2} = \int \left( \frac{1}{2} \left( (D_t \xi)^2 + (D_t \eta)^2\right) - (f_0 + \beta y) \eta D_t \xi - \frac{\beta U}{2} \eta^2 + q \nabla \cdot \bxi - \frac{1}{2} \bxi \cdot \nabla \nabla P \cdot \bxi \right) \, \d \bx,
\eeq
where $D_t=\partial_t + \bs{U} \cdot \nabla$, with $\bs{U}$ the background velocity satisfying $\nabla \cdot \bs{U} = 0$. The background-flow positions $\bs{X}$ are now denoted by $\bs{x}$ and  used as independent variables. 
Taking the variations of $ \int L\su{2} \, \d t$ with respect to $\bxi$ and $q$ yields linearised equations for the perturbation to the background flow. 

The main balance for inertial waves is given by
\beq \lab{iwbal}
\partial_t \xi - f_0 \eta = 0, \quad \partial_t \eta + f_0 \xi = 0, \quad \nabla \cdot \bxi = 0.
\eeq
To study modulations of such waves, we follow YBJ and let
\beq \lab{M}
\xi + \i \eta = M_z \e^{-\i f_0 t}, \quad \zeta =-  M_\chi \e^{-\i f_0 t} + \cc,
\eeq
where $M$ is a complex amplitude and $\chi =x+\i y$ (so that $\partial_\chi=(\partial_x-\i\partial_y)/2$). These satisfy \eqn{iwbal} identically for time-independent $M$. Whitham's prescription to obtain an amplitude equation for $M$ is to introduce \eqn{M} into \eqn{L2}, average over the wave phase $f_0 t$, and regard the resulting Lagrangian as a functional of $M$ that is stationary \citep{whit74}. 

We now follow this prescription. Using $\langle \cdot \rangle$ to denote phase average, we compute
\beq \lab{kin}
\av{(D_t \xi)^2 + (D_t \eta)^2} = \left| D_t M_z \right|^2 - \i f_0 \left( M_z  D_t M_z^*-  M_z^* D_t M_z\right) + f_0^2 |M_z|^2,
\eeq
\beq \lab{cor}
\langle \eta D_t \xi \rangle = -\frac{\i}{4} \left( M_z  D_t M_z^*-  M_z^* D_t M_z\right) + \frac{f_0}{2} |M_z|^2, \quad \langle \eta^2 \rangle = \frac{1}{2} |M_z|^2.
\eeq
and 
\beq \lab{press}
\langle \bxi \cdot \nabla \nabla P \cdot \bxi \rangle = 2 \left( P_{\chi \chi^*} |M_z|^2 + P_{zz} |M_\chi|^2 - P_{\chi z} M_z M_\xi^* - P_{\chi^* z} M_z^* M_\xi\right).
\eeq
Introducing \eqn{kin}--\eqn{press} into \eqn{L2} and neglecting terms not proportional to $f_0$ in the first four terms we finally obtain the averaged Lagrangian
\begin{eqnarray*}\lab{Lav}
\bar L &=& \int \left( \frac{-\i f_0}{4} \left( M_z  D_t M_z^*-  M_z^* D_t M_z\right) - \frac{f_0 \beta y}{2} |M_z|^2  \right. \\
&& \qquad   - \left( P_{\chi \chi^*} |M_z|^2 + P_{zz} |M_\chi|^2 - P_{\chi z} M_z M_\chi^* - P_{\chi^* z} M_z^* M_\chi\right)  \bigg) \, \d \bx.
\end{eqnarray*}
Scaling this expression by $\i f_0/2$ and integrating by parts, we obtain
\begin{eqnarray*}\lab{Lav}
\bar L &=& \int \bigg( M_z D_t M_z^* - \i \beta y |M_z|^2   \\
&& \qquad  \left. - \frac{2\i}{f_0}\left( P_{\chi \chi^*} |M_z|^2 + P_{zz} |M_\chi|^2 - P_{\chi z} M_z M_\chi^* - P_{\chi^* z} M_z^* M_\chi\right) \right) \, \d \bx.
\end{eqnarray*}
This can be written in terms of $\bx$ and further simplified by noting that
\begin{eqnarray*}
P_{zz} |M_\chi|^2 - P_{\chi z} M_z M_\chi^* - P_{\chi^* z} M_z^* M_\chi &=& \tfrac{1}{4} \left(P_{zz} |\nabla M|^2 - \nabla P_z \cdot \left(\nabla M^* M_z + \nabla M M^*_z\right) \right) \\ && \qquad + \i \nabla \cdot (\cdots),
\end{eqnarray*}
where, from now on, $\nabla=(\partial_x,\partial_y)$.
This reduces the Lagrangian to its final form
\begin{eqnarray}\lab{Lav}
\bar L &=& \int \bigg( M_z D_t M_z^* - \i \beta y |M_z|^2  \nonumber \\
&& \qquad  \left. - \frac{\i}{2f_0}\left( \nabla^2 P |M_z|^2 + P_{zz} |\nabla M|^2 - \nabla P_z \cdot \left(\nabla M^* M_z + \nabla M M^*_z\right) \right) \right) \, \d \bx.
\end{eqnarray}
Taking the variation with respect to $M^*$ gives, after simplifications,
\beq \lab{YBJ}
(D_t M_z)_z + \i \beta y M_{zz} + \frac{\i}{2f_0} \left(\nabla^2 P M_{zz} + P_{zz} \nabla^2 M - 2 \nabla P_z \cdot \nabla M_z \right)=0.
\eeq
This is a slight generalisation of Eq.\ (3.2) in YBJ, with the form (3.4a) of their bracket. Our derivation makes no assumption of geostrophic balance for the background flow which can be fully three dimensional. In contrast, \citet{youn-benj}  obtained their equation making this assumption: using the consequences $\bs{U} = (\nabla^\bot \Psi,0)$, $\nabla P = f_0 \nabla \Psi$ and $P_{zz} = N^2 + f_0 \Psi_{zz}$, where $\Psi$ is the mean-flow streamfunction and $N$ is the Brunt--V\"ais\"al\"a frequency, reduces \eqn{YBJ} to their  Eq.\ (3.2).
Note that improved approximations can be derived by retaining more terms in the horizontal kinetic energy, namely the term $|D_t M_z|^2$, the vertical component of the kinetic energy, and the non-traditional part omitted from the Coriolis term. 

The Hamiltonian form of \eqn{YBJ} is found by inspection. Let $\mathcal{U}=M_z$, then we can rewrite \eqn{YBJ} (with $\beta=0$ for simplicity) as
\[
\i \mathcal{U}_t = - \i \bs{U} \cdot \nabla \mathcal{U} + \frac{1}{2 f_0} \int \left(\nabla^2 P \mathcal{U}_{z} + P_{zz} \nabla^2 \int \mathcal{U} \, \d z - 2 \nabla P_z \cdot \nabla \mathcal{U} \right) \, \d z= H \mathcal{U},
\]
where the second equality defines the Hamiltonian operator. This can be shown to be Hermitian, $H^*=H$, by writing it in the form (3.4b) in YBJ, namely
\[
H \mathcal{U} = - \i \bs{U} \cdot \nabla \mathcal{U} + \frac{1}{2 f_0} \left( \int \nabla^2 (P \mathcal{U}_z) \, \d z - 2 \nabla \cdot ( P \cdot \nabla \mathcal{U}) + \left(P \nabla^2 \int \mathcal{U} \, \d z \right)_z \right).
\]
It follows from the Hamiltonian form of the equation that
\[
\mathcal{A}=\tfrac{1}{2} \int |\mathcal{U}|^2 \, \d \bs{x} \inter{and} \mathcal{H} = \tfrac{1}{2} \int \mathcal{U}^* H \mathcal{U} \, \d \bs{x}
\]
are conserved:
\[
\dt{\mathcal{A}}{t} = -\frac{\i}{2} \int (\mathcal{U}^* H \mathcal{U} - \mathcal{U}^* H^* \mathcal{U}) \, \d \bs{x} = 0
\]
and
\[
\dt{\mathcal{H}}{t} = -\frac{\i}{2} \int (\mathcal{U}^* H H \mathcal{U} - \mathcal{U}^* H^* H \mathcal{U}) \, \d \bs{x} = 0.
\]
The first conservation law, obtained in YBJ, stems from the independence of the Hamiltonian on the phase of $\mathcal{U}$ (symmetry $\mathcal{U} \mapsto \mathcal{U} \exp(\i \phi)$);
the second, which was pointed out by B\"uhler (personal communication), stems from the time independence of the Hamiltonian. This justifies the terminology of action for $\mathcal{A}$ and energy for $\mathcal{H}$.

YBJ also derived a simplified model, their Eq.\ (4.7), assuming that the vertical scale of the mean flow is much larger than that of the waves. This simplified model is derived straightforwardly from the Lagrangian \eqn{Lav} by omitting the last two terms (proportional to $\nabla P_z$) and approximating $P_{zz}$ by $N^2$ to obtain
\[
\bar L = \int \left( M_z D_t M_z^* - \i \beta y |M_z|^2   - \frac{\i}{2f_0}\left( \nabla^2 P |M_z|^2 + N^2 |\nabla M|^2 \right) \right) \, \d \bs{x}. 
\]
Taking variations with respect to $M^*$ yields 
\[
(D_T  M_z)_z + \i \beta y M_{zz} + \frac{\i N^2}{2 f_0} \nabla^2 M + \frac{\i}{2 f_0} ( \nabla^2 P M_z)_z  = 0,
\]
which can be recognised as YBJ's Eq.\ (4.7) when $\nabla P = f_0 \nabla \Psi$. Rewriting the Lagrangian in terms of $\mathcal{U}=M_z$
gives another version,
\[
\bar L = \int \left( \mathcal{U} D_t \mathcal{U}^*  - \i \beta y |\mathcal{U} |^2   - \frac{\i f_0}{2} \mathcal{U}^* L^{-1} \nabla^2 \mathcal{U}- \frac{\i}{2f_0}\nabla^2 P |\mathcal{U}|^2  \right) \, \d \bs{x},
\]
where the linear operator $L$ is defined by
\[
L A = (f_0^2 N^{-2} A_z)_z.
\]
Variations with respect to $\mathcal{U}$ give YBJ's (4.5) with $\mathcal{U}=L A$. This model conserves the same action as the more general YBJ model, and the energy
\[
\mathcal{H} = \frac{1}{2} \int \left( - \i \mathcal{U}^* \bs{U} \cdot \nabla \mathcal{U} +  \beta y |\mathcal{U} |^2   + \frac{f_0}{2} \mathcal{U}^* L^{-1} \nabla^2 \mathcal{U} + \frac{1}{2f_0}\nabla^2 P |\mathcal{U}|^2  \right) \, \d \bs{x}.
\]
\smallskip

\noindent
\textbf{Acknowledgements.} The author thanks O. B\"uhler, E. Danioux,  W. R. Young and J.-H. Xie for valuable discussions. This research is funded by the UK Natural Environment Research Council (grant NE/J022012/1). 

\bibliographystyle{plainnat}
\bibliography{mybib}

\end{document}